\documentstyle[prl,aps,twocolum,psfig]{revtex}

\begin{document}

\hoffset=-0.5in
\textwidth=7in
\preprint{LBNL-41741, DOE/ER/40561-5-INT98}

\twocolumn[\hsize\textwidth\columnwidth\hsize\csname @twocolumnfalse\endcsname
\title{Where is the jet quenching in $Pb+Pb$ collisions at 158 AGeV?}
\author{Xin-Nian Wang}
\address{
Nuclear Science Division, Mailstop 70A-3307,\\
Lawrence Berkeley National Laboratory, Berkeley, CA 94720 USA\\
and\\
Institute for Nuclear Theory, University of Washington\\
Seattle, WA 98195-1550}
\date{April 20, 1998}
\maketitle

\begin{abstract}

Because of the rapidly falling particle spectrum at large $p_T$ from jet 
fragmentation at the CERN SPS energy, the high-$p_T$ hadron distribution 
should be highly sensitive to parton energy loss inside
a dense medium as predicted by recent perturbative QCD (pQCD)
studies. A careful analysis of recent data from CERN SPS experiments via
pQCD calculation shows little evidence of energy loss. This implies
that either the life-time of the dense partonic matter is very short 
or one has to re-think about the problem of parton energy loss in
dense matter. The hadronic matter does not seem to cause jet quenching 
in $Pb+Pb$ collisions at the CERN SPS. High-$p_T$ two particle correlation 
in the azimuthal angle is proposed to further clarify this issue.
\end{abstract}
\pacs{25.75.+r, 12.38.Mh, 13.87.Ce, 24.85.+p}
 ]

Hard processes have been considered good probes of the dense matter
which is produced in high-energy heavy-ion collisions and is expected
to be in the form of deconfined quarks and gluons or a quark-gluon
plasma (QGP) at high energy densities. These processes happen in the
earliest stage of the collisions and therefore can probe the
properties of the dense matter in its early form, whether a QGP or
not. Furthermore, their production rates can be calculated with
reasonable accuracy within pQCD parton model and has been tested
extensively against vast experimental data in $p+p$ and $p+A$
collisions. These calculations \cite{HPC} incorporating minimum amount
of normal nuclear effects (nuclear modification of parton distributions
\cite{EMC} and Cronin effect \cite{cronin-ex1}) then provide a clean
and reliable baseline against which one can extract signals of the
dense matter. In this paper, we investigate what high-$p_T$ particles 
from jet fragmentation tell us about the dense matter formed in
$Pb+Pb$ collisions at the CERN SPS.

        Like other hard processes, large transverse momentum parton
jets are produced in the early stage of high-energy heavy-ion
collisions. They often have to travel through the dense matter
produced in the collisions and finally hadronize into high-$p_T$
particles in the central rapidity region. Recent theoretical studies
\cite{GW1,BDPS,BDMPS,BGZ} show that a fast parton will lose a
significant amount of energy via induced pQCD radiation when it
propagates through a dense partonic matter where the so-called 
Landau-Pomeranchuk-Migdal coherence effect becomes important.
If this picture of parton energy loss can be applied to large 
transverse momentum parton jets in the central rapidity region of 
high-energy central $A+A$ 
collisions, one should expect a leading parton to lose energy 
when it propagates through a long-lived dense matter.
Since the radiated gluons will eventually become incoherent from
the leading parton which will fragment into large-$p_T$ hadrons, one
then should expect a reduction of the leading hadron's $p_T$ or a
suppression of the large-$p_T$ particle
spectrum \cite{WG92,WH,wang98}. At the CERN SPS energy, high-$p_T$
jet or particle
production ($p_T > 3 $ GeV/$c$) is very rare and the power-law-like
spectrum is very steep because of the limited phase space. It should
be especially sensitive to any finite energy loss.

The single inclusive particle spectrum at large $p_T$ in high-energy
$p+p$ or $p+\bar p$ collisions can be calculated in a pQCD parton
model with the information of parton distributions \cite{mrs} and jet
fragmentation functions \cite{bkk} from deep-inelastic $e+p$ and
$e^+e^-$ experiments. This is one of the early successes of the QCD
parton model \cite{feynman,owens2,owens}. It was already pointed out 
that the initial transverse momentum before the hard scattering
is very important to take into account at lower energies and can
significantly increase the single inclusive differential cross
section. The initial parton transverse momentum can be studied in
detail via Drell-Yan (DY) \cite{dy,altarelli,collins,field}, $\gamma+$jet
and $\gamma+\gamma$ production in $p+p$ collisions.

To the lowest order of pQCD, the single inclusive particle  production
cross section can be written as \cite{owens},
\begin{eqnarray}
  \frac{d\sigma_h^{pp}}{dyd^2p_T}&=&K\sum_{abcd}
  \int dx_a dx_b d^2k_{aT} d^2k_{bT} g_p(k_{aT},Q^2) 
  \nonumber \\ & & g_p(k_{bT},Q^2) f_{a/p}(x_a,Q^2)f_{b/p}(x_b,Q^2)
   \nonumber \\ & & 
  \frac{D^0_{h/c}(z_c,Q^2)}{\pi z_c}
  \frac{d\sigma}{d\hat{t}}(ab\rightarrow cd), \label{eq:nch_pp}
\end{eqnarray}
where $x_{a,b}$ are the fractional energies and $k_{a,bT}$ the initial
transverse momenta of the colliding partons. 
$d\sigma/d\hat{t}(ab\rightarrow cd)$ are the differential
elementary parton-parton cross sections \cite{owens}.  $K\approx 2$ is used to 
account for higher order corrections \cite{xwke} 
and $Q=P_{cT}=p_T/z_c$. We will use MRSD$-'$ parameterization for
the parton distributions $f_{a/p}(x,Q^2)$ and BKK parameterization for
the jet fragmentation functions $D^0_{h/c}(z,Q^2)$. We will use a
Gaussian form for the initial-$k_T$ distribution 
$g_p(k_T,Q^2)=1/(\pi\langle k_T^2\rangle_p) 
\exp(-k_T^2/\langle k_T^2\rangle_p)$ with a
variance $\langle k_T^2\rangle_p=$ 1 (GeV$^2/c^2$) + $0.2
Q^2\alpha_s(Q^2)$, where the $Q$-dependence accounts for initial $k_T$
from initial-state radiation ( or higher order $2\rightarrow 2+n$
processes) \cite{field}. The parameters are chosen to best fit the
experimental data of high-$p_T$ particle spectra at all energies
\cite{wangsps}. Because of the introduction of initial parton $k_T$,
one of the Mandelstam variables for the elementary parton-parton
scattering processes could vanish and cause the differential parton
cross sections to diverge in certain phase space points. We use an effective
parton mass $\mu=0.8$ GeV to regulate the divergence as in the early
studies \cite{feynman}. The resultant spectrum is sensitive to the
value of $\mu$ only at around $p_T\sim \mu$, where pQCD calculation
is not reliable in any case.

\begin{figure}
\centerline{\psfig{figure=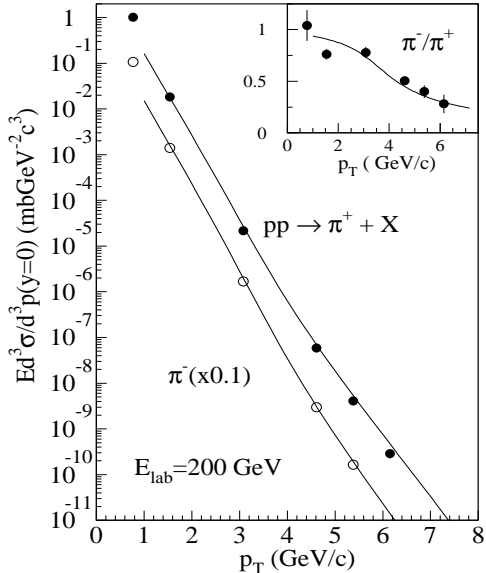,width=2.5 in,height=3.0in}}
\caption{ Single-inclusive pion spectra in $p+p$ collisions at 
$\protect E_{\rm lab}=200$ GeV. The solid lines are from 
pQCD calculations and data
from Ref.~\protect\cite{cronin-ex1}. The inserted figure are the corresponding
$\protect\pi^-/\pi^+$ ratios.}
\label{figlet1}
\end{figure}

        Shown in Fig.~\ref{figlet1} is an example of the calculated
$\pi^{\pm}$ spectra in $p+p$ collisions at $E_{\rm lab}=200$ GeV. The
agreement with experimental data is very good not only for the overall
inclusive cross section but also for the iso-spin dependence as shown
by the $p_T$-dependence of the $\pi^-/\pi^+$ ratio in the inserted
figure. Similar analyses have been carried out at other energies up to
Fermilab Tevatron \cite{wangsps}. The initial $k_T$ is less important
and becomes almost negligible for the single inclusive parton spectra
at these collider energies.

        In $p+A$ collisions, there are two known nuclear
effects: nuclear modification of the parton distributions 
(EMC effect) \cite{EMC} and nuclear enhancement of the large-$p_T$ hadron 
spectra (Cronin effect) \cite{cronin-ex1}. Both are caused
by multiple initial scattering. We assume that the parton
distributions per nucleon inside a nucleus at impact parameter $b$,
\begin{eqnarray}
        f_{a/A}(x,Q^2,b)&=&S_{a/A}(x,b)\left[ \frac{Z}{A}f_{a/p}(x,Q^2) \right.
        \nonumber \\
        &+&\left. (1-\frac{Z}{A}) f_{a/n}(x,Q^2)\right],
\end{eqnarray}
is factorizable into the parton distributions inside a normal nucleon
and the nuclear modification factor, $S_{a/A}(x,b)$, for which we use the
HIJING parameterization \cite{hijing}. This should be adequate at the
CERN SPS energy where the dominant process at large $p_T$ is
quark-quark scattering.

One can explain the Cronin effect within a multiple parton scattering 
model\cite{wang-rep,cronin-th1}, in which the cancellation by the 
absorptive processes forces the nuclear enhancement to disappear at 
large $p_T$ like $1/p_T^2$ and in the meantime causes a slight suppression of
hadron spectra at small $p_T$ so that the integrated spectra do not
change much . This allows us to take into account the effect of multiple 
scattering via a broadening of the initial transverse momentum, 
\begin{equation}
\langle k_T^2\rangle_A(b)= \langle k_T^2\rangle_p+[\nu(b)-1]\Delta^2,
\label{eq:wda}
\end{equation}
where $\nu(b)=\sigma_{pp}t_A(b)$ is the average number of scattering
the parton's parent nucleon has suffered and $t_A(b)$ is the nuclear
thickness function normalized to $\int d^2b t_A(b)=A$. Since the Gaussian
distribution is not a good approximation for the $k_T$-kick during 
the initial multiple scattering, we found that we have to use a 
scale-dependent value,
$\Delta^2=0.225\ln^2(Q/{\rm GeV})/(1+\ln(Q/{\rm GeV}))$ GeV$^2/c^2$,
to best describe the available data from $p+A$ collisions
\cite{wangsps} which allow about $10-20\%$ uncertainty in the
calculated spectra. For $Q=2\sim 3$ GeV, 
$\Delta^2=0.064 \sim 0.129$ GeV$^2/c^2$, which is 
consistent with the analyses of $p_T$ broadening for $J/\Psi$ production 
in $p+A$ \cite{gg,dns2}.

Taking into account these nuclear effects which already exist
in $p+A$ collisions, the single inclusive particle spectra in $A+A$
collisions can be estimated as
\begin{eqnarray}
  \frac{d\sigma_h^{AA}}{dyd^2p_T}&=&K\sum_{abcd} \int d^2b \int d^2r
  t_A(r)t_A(|{\bf b}-{\bf r}|)
  \int dx_a dx_b \nonumber \\
  & & d^2k_{aT} d^2k_{bT} 
  g_A(k_{aT},Q^2,r) g_A(k_{bT},Q^2,|{\bf b}-{\bf r}|)  \nonumber \\
  & & f_{a/A}(x_a,Q^2,r) f_{b/A}(x_b,Q^2,|{\bf b}-{\bf r}|) \nonumber \\
  & & \frac{D^0_{h/c}(z_c,Q^2)}{\pi z_c}
  \frac{d\sigma}{d\hat{t}}(ab\rightarrow cd). \label{eq:nch_AA}
\end{eqnarray}
The initial-$k_T$ distribution $g_A(k_T,Q^2,b)$ is similar to that of
a proton in Eq.~(\ref{eq:nch_pp}) with a broadened width given by
Eq.~(\ref{eq:wda}) which now depends on the impact-parameter $b$.

For central $A+A$ collisions, we limit the integration over the impact
parameter to $b_{\rm max}$. Using the geometrical cross section of a
hard-sphere nucleus, we determine $b_{\rm max}$ by matching $b_{\rm
max}^2/4\pi R_A^2$ ($R_A\approx 1.12A^{1/3}$ fm) 
to the fractional cross section
of the triggered central events in experiments. In
Eq.~\ref{eq:nch_AA}, we actually use the Wood-Saxon distribution to
calculate the thickness function $t_A(b)$.

\begin{figure}
\centerline{\psfig{figure=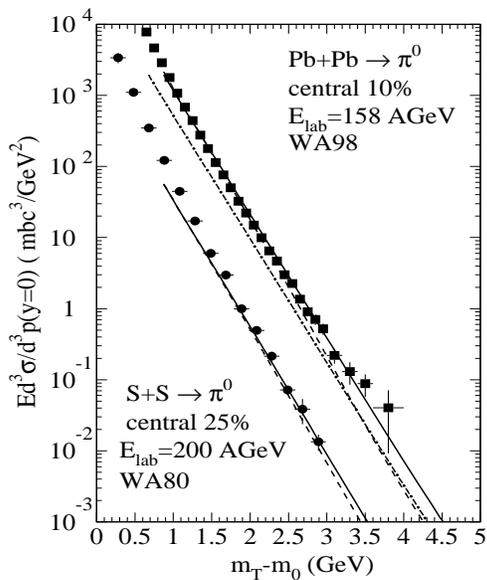,width=2.5in,height=3.0in}}
\caption{ Single-inclusive $\pi^0$ spectra in central $S+S$ at 
$\protect E_{\rm lab}=200$ GeV and $Pb+Pb$ collisions at $E_{\rm lab}=158$
GeV.  The solid lines are pQCD calculations with initial-$k_T$
broadening and dashed lines are without. The $S+S$ data are from WA80
\protect\cite{wa80} and $Pb+Pb$ data are from WA98 
\protect\cite{wa98}. The dot-dashed line is obtained from the solid line
for $Pb+Pb$ by shifting $p_T$ by 0.2 GeV/$c$.}
\label{figlet2}
\end{figure}

Shown in Fig.~\ref{figlet2} are the calculated single-inclusive
spectra for $\pi^0$ in central $S+S$ ($E_{\rm lab}=200$ GeV) and
$Pb+Pb$ ($E_{\rm lab}=158$ GeV) collisions with (solid) and without
(dashed) nuclear $k_T$-broadening as compared to WA80
\cite{wa80} and WA98 \cite{wa98} data. 
Besides small effects of the nuclear modification of the parton 
distributions on the spectra at these energies,
the dashed lines are simply the spectra in $p+p$
collisions multiplied by the nuclear geometrical factor. It is clear
that one has to include the $k_T$-broadening due to the initial
multiple scattering in order to describe the data. This is also
consistent with the analysis by WA80 \cite{wa80}.

One can conclude from this analysis that the factorized pQCD parton model
seems to work well for large-$p_T$ hadron production in $A+A$
collisions. But one can also immediately realize that there is no
evidence of parton energy loss as predicted by previous theoretical studies
\cite{GW1,BDPS,BDMPS,BGZ}. If there is parton energy loss and the
radiated gluons become incoherent from the leading parton, the effective
fragmentation functions should be modified such that the leading
high-$p_T$ particles should be suppressed as compared to $p+p$ and
$p+A$ collisions \cite{WG92,WH,wang98}. At $y=0$, parton energy loss
can be directly translated into $p_T$ reduction for the leading
hadrons. To estimate the experimental constraints on parton
energy loss, one can simply shift the $p_T$ values of the solid line
for $Pb+Pb$ in Fig.~\ref{figlet2} by 0.2 GeV/$c$ (dot-dashed
line). Assuming $20\%$ uncertainty of the calculated spectrum, one can
quickly exclude a total energy loss $\Delta E<0.1$ GeV. With the
transverse size of a $Pb$ nucleus, this corresponds to an energy loss
$dE/dx<0.02 $ GeV/fm. Detailed model calculations will give a more
stringent limit \cite{wangsps}. This is in direct contradiction with 
the current theoretical studies of parton
energy loss in dense matter and calls into question current 
models of energy loss. It also implies that 
there is not a dense partonic matter which exists long enough to cause 
parton energy loss.

        Most of the recent theoretical studies
\cite{GW1,BDPS,BDMPS,BGZ} of energy loss are based on pQCD calculation
for a single fast parton propagating through a large dense medium.
If we assume that it is valid for a parton
propagating through a deconfined medium, the absence of parton energy
loss in the experimental data on high-$p_T$ particle spectra implies 
that either there is no such deconfined partonic matter being formed or it
only lived for a very short period of time.
Using the measured $dE_T/d\eta \approx 405 $ GeV
\cite{na49-et} in the central rapidity region of most central
$Pb+Pb$ collisions (\%2 of the total inelastic cross section) one can
estimate the initial energy density at $\tau_0=1$ fm/$c$ to be about
$\epsilon_0=dE_T/d\eta/(\pi \tau_0 R_A^2) \approx 2.9 $ GeV/fm$^3$.
This is an optimistic estimate assuming that the formation time of the 
dense matter is about 1 fm/$c$. 
Because of longitudinal expansion, the energy density will 
decrease like $\epsilon/\epsilon_0=(\tau_0/\tau)^{\alpha}$. The value
of $\alpha$ could range from 1 for free-streaming to 4/3 for
hydro-expansion of an ideal gas of massless particles. Assuming a
critical energy density of $\epsilon_c\approx 1$ GeV/fm$^3$, the
system can only live above this critical density for about $2.2\sim 2.9$
fm/$c$. Equilibrating processes and transverse expansion
certainly will reduce this life-time even further. During such a short
time, a highly virtual parton has small interaction cross section
before its virtually decreases through pQCD evolution.
Therefore, a
produced large $p_T$ parton will not have much time to lose
its energy before the dense matter drops below the critical
density. 
The recent theoretical studies \cite{GW1,BDPS,BDMPS,BGZ} are not applicable to 
such a short-lived system. Nevertheless, this analysis at least 
tells us that the life-time 
of the dense partonic matter must be short if it is ever formed 
in $Pb+Pb$ collisions at 158 AGeV. 
Otherwise, it is difficult to reconcile the absence of parton energy 
loss with the strong parton interaction which drives the equilibration and
maintains a long life-time of the initial parton system.

        One definite conclusion one can draw from this analysis is
that the hadronic matter in the later stage of heavy-ion collisions 
does not seem to cause parton energy loss or jet quenching at the
CERN SPS. This will make jet quenching an even better probe of 
long-lived initial partonic matter since it will not be affected by 
the hadronic phase of the matter.
Because of its long formation time ($\tau_f\sim 20$ fm/$c$
for a pion with $p_T\sim 3$ GeV/$c$), a high-$p_T$ pion is
only formed either after freeze-out or in a very dilute hadronic matter.
Otherwise, inelastic scattering with other soft pions can also cause
the suppression of high-$p_T$ particle spectra or apparent jet
quenching. What is traveling through the hadronic matter is thus a
fragmenting parton whose interaction with a hadronic matter might be
non-perturbative in nature. The pQCD estimate of parton energy loss is
then not applicable here even though it might be adequate for a parton
propagating in a hot QGP. The fact that a fragmenting parton does
not lose much energy in hadronic matter might be related to the
absence of parton energy loss
to the quarks and anti-quarks prior to DY hard processes in $p+A$
and $A+A$ collisions.

The initial energy density at RHIC is expected to be higher than at SPS.
If one observes significant suppression of large-$p_T$ hadrons at RHIC as 
was predicted \cite{WG92,WH,wang98}, it clearly reveals an initial condition 
dramatically different from the CERN SPS.

The observed high-$p_T$ pion spectra in central $Pb+Pb$ collisions
cannot be due to collective hydrodynamic flow, since there will always
be high-$p_T$ partons produced in the coronal region of the two
overlapped nuclei where jet propagation and fragmentation will not be
influenced by the dense matter. To verify that these spectra are
from jet production and fragmentation rather than from hydrodynamic flow,
one can measure the azimuthal particle correlation (selecting
particles above a certain $p_T$) relative to a triggered
high-$p_T$ particle as was proposed in Ref.~\cite{wangcorr}. One
should see a double-peak structure characteristic of a jet profile.
One can use this method at even moderate $p_T$ (where there are still
not many particles per event) to determine the contribution 
from semi-hard processes and the $p_T$ range for which use of a thermal 
fire-ball model is justified. Otherwise, the extracted temperature and 
radial flow velocity can be misleading.

\section*{Acknowledgments}
The author would like to thank M. Gyulassy, U. Heinz, B. Jacak 
and B. M\"uller for intense
discussions on the implications of the absence of energy loss at the
CERN SPS. He would also like to thank T. Peitzmann for communication
regarding WA80 and WA98 data. This work was supported by DOE under 
Contract No. DE-AC03-76SF00098.
The author wishes to thank the
Institute for Nuclear Theory for kind hospitality during his stay when
this work was written.

\end{document}